\documentclass[prb,aps,twocolumn]{revtex4}
\usepackage{epsfig}
\usepackage{amsmath}
\draft
\begin{document}
\title{Phase diagram of model anisotropic particles with octahedral symmetry}
\author{E. G. Noya}
\author{C. Vega} 
\affiliation{Departamento de Qu\'{\i}mica-F\'{\i}sica, Facultad de
	Ciencias Qu\'{\i}micas, Universidad Complutense de Madrid,
	E-28040 Madrid, Spain}
\author{J. P. K. Doye} 
\affiliation{Physical and Theoretical Chemistry Laboratory, Oxford
	University, South Parks Road, Oxford, UK OX1 3QZ}
\author{A. A. Louis} 
\affiliation{Rudolf Peierls Centre for Theoretical Physics, 1 Keble
	Road, Oxford, UK OX1 3NP}
\date{\today}

\begin{abstract}

We computed the phase diagram for a system of model anisotropic particles
with six attractive patches in an octahedral arrangement.
We chose to study this model for a relatively narrow value of the patch width
where the lowest-energy configuration of the system is a simple cubic crystal.
At this value of the patch width, there is no stable vapour-liquid phase 
separation, and there are three other crystalline phases
in addition to the simple cubic crystal that is most stable at low pressure.
Firstly, at moderate pressures, it is more favourable
to form a body-centred cubic crystal, which can be viewed as two 
interpenetrating, and almost non-interacting, simple cubic lattices.
Secondly, at high pressures and low temperatures,
an orientationally ordered face-centred cubic structure becomes
favourable.
Finally, at high temperatures a face-centred cubic plastic 
crystal is the most stable solid phase.
\end{abstract}

%\pacs{61.46.+w,36.40.Ei,36.40.Mr}
%36.40.Ei Phase transitions in clusters
%36.40.Mr Spectroscopy and geometrical structure of clusters
%61.46.+w Nanoscale materials: clusters, nanoparticles, nanotubes, 
%and nanocrystals (see also 36.40.-c
%Atomic and molecular clusters; for fabrication and characterization of nanoscale materials, see 81.07.-b in
%materials science)
%64.70.Dv Solid-liquid transitions

\maketitle

\vspace{0.5cm}
%}
%%%%%%%%%%%%%%%%%%%%%%%%%%%%%%%%%%%%%%%%%%%%%%%%%%%%%%%%%%%%%%%%%%%%%%%
%%%%%%%%%%%%%%%%%%%%%%%%%%%%%%%%%%%%%%%%%%%%%%%%%%%%%%%%%%%%%%%%%%%%%%%
\section{Introduction}
%%%%%%%%%%%%%%%%%%%%%%%%%%%%%%%%%%%%%%%%%%%%%%%%%%%%%%%%%%%%%%%%%%%%%%%

	Anisotropic patchy particles consisting of a repulsive core with
some attractive sites have been used to study a variety of  
problems. The first anistropic patchy potentials were introduced 
as models of associative liquids (see, for example, Refs.
\onlinecite{kolafa,jackson_molphys1988,sear_jcp1996,vegamonson,galindo_molphys1998,monson_jcp2004}).
More recently, anisotropic patchy models have received renewed interest in the
context of protein 
crystallization.\cite{sear_jcp1999,song_pre2002,dixit_jcp2002,frenkel_jcp2003,sandler_jcp2004,sandler_jpcb2005,talanquer_jcp2005}
Proteins are usually hard to crystallize, but, in order
to determine the structure of a given protein, and hence its functionality,
large crystals that can be used in high-resolution X-ray diffraction studies 
are needed.\cite{vekilov}
In this respect, theoretical studies aiming to predict the
conditions which favour crystallization would be very valuable.
Even though some significant progress has already been
made using simple isotropic potentials,\cite{rosenbaum,tenwolde}
it is known that protein interactions are short-ranged
and highly anisotropic. For example, most protein crystals have packing
fractions that are much lower than the close-packed solids typically favoured
by isotropic potentials.\cite{Matthews68}
So far, studies using anisotropic models have shown that the fluid-fluid coexistence moves to lower temperatures as the interactions become more anisotropic 
(either by decreasing the number of patches or making them smaller)\cite{frenkel_jcp2003}
and can even become metastable.\cite{sear_jcp1999} The introduction
of anisotropy can also induce the stability of multiple solid phases, 
including orientationally ordered and plastic phases.\cite{sandler_jcp2004}

Anisotropic interactions have also received much attention
in the context of the development of new materials.
In particular, there has been increasing interest in the fabrication of 
colloids\cite{vanderhoff,pine_sicence2003,pine_jacs2005,pine_chemmat2005,Roh05,Snyder05,Li05b,blaaderen_science2006,Roh06} and nanoparticles\cite{Jackson04,Levy06b,Devries07} with anisotropic interactions, one of the goals
being to tailor their interactions so that they are able to self-assemble
into a given target structure. One set of targets that has 
received particular attention is low-density colloidal crystals, 
e.g.\ colloidal diamond, because of their potential as photonic materials.\cite{Hynninen07}
This work has stimulated a number of recent theoretical studies that have begun
to address the question of how such patchy interactions can be used to control
the crystallization behaviour\cite{glotzer_langmuir2005,jon} 
and the self-assembly of finite objects of given size and symmetry.\cite{glotzer_nanoletters2004,chandler,vanworkum,alex} 
Some of the latter studies have also been motivated by a desire to understand
biological self assembly, such as the formation of virus capsids.

The final area where anisotropic patchy models have been the subject of 
recent interest is in the study of the dynamics of 
supercooled liquids.\cite{demichelle,bianchi,sciortino_arxiv2007,sciortino_pmw} 
In particular, as patchy models tend to move the liquid-vapour coexistence
line to lower temperature, and to lower packing fractions,\cite{bianchi} 
they make it much easier to study gels, 
i.e.\ dynamically-arrested states that have low density and 
where the arrest is due to formation of energetically stable bonds, 
rather than caging of the atoms in a densely-packed environment.

In spite of these studies, there is still much to be learnt from 
a fundamental point of view about how anisotropic interactions 
affect the thermodynamics and dynamics of a system. 
In particular, it is not fully understood how the arrangement of 
the patches will affect the phase diagram. It seems reasonable that if 
the patches are located at positions that favour the local environment
of a given crystalline structure, crystallization into that structure would
be favoured. On the other hand, a random distribution of the patches, as 
perhaps may be the case for the surface of a protein,
will normally lead to a situation where the local order is
not compatible with any crystalline lattice and, therefore,
crystallization will be hindered. 
However, relatively little is known about how the 
specificity of the angular interactions will affect the 
relative stability of the phases
or their accessibility due to kinetic effects.  

In this paper we study a model of patchy particles that we have previously
used to study the self-assembly of monodisperse clusters,\cite{alex} 
and the kinetics of crystallization in two and three dimensions.\cite{jon}
One of the intriguing results of the latter work was that crystallization
of particles with six octahedrally-arranged patches into a simple cubic 
(sc) crystal appeared to be much easier than crystallization of particles with
four tetrahedrally-arranged patches into a diamond lattice. Our hypothesis
is that this difference in crystallization kinetics reflects the absence of 
frustration in the octahedral system, whereas in the tetrahedral system
the preferred local order differs from the global crystalline order, thus 
frustrating crystallization. In order to understand this further a systematic 
study of the nucleation behaviour for these two systems is required, and a 
necessary precursor for such work is the computation of the phase diagram.
This is one of the motivations for the present study, where we compute
the phase diagram for particles with an octahedral arrangement of the patches.
Furthermore, the phase diagram will also be of considerable interest in its own
right, with the potential for competition between lower- and higher-density 
crystalline phases, and the possibility of plastic phases that have 
translational, but no orientational order.

%%%%%%%%%%%%%%%%%%%%%%%%%%%%%%%%%%%%%%%%%%%%%%%%%%%%%%%%%%%%%%%%%%%%%%%
%%%%%%%%%%%%%%%%%%%%%%%%%%%%%%%%%%%%%%%%%%%%%%%%%%%%%%%%%%%%%%%%%%%%%%%
\section{Method}
\label{method}
%%%%%%%%%%%%%%%%%%%%%%%%%%%%%%%%%%%%%%%%%%%%%%%%%%%%%%%%%%%%%%%%%%%%%%%

%%%%%%%%%%%%%%%%%%%%%%%%%%%%%%%%%%%%%%%%%%%%%%%%%%%%%%%%%%%%%%%%%%%%%%%
\subsection{Model}
%%%%%%%%%%%%%%%%%%%%%%%%%%%%%%%%%%%%%%%%%%%%%%%%%%%%%%%%%%%%%%%%%%%%%%%

\begin{figure}[!t]
\begin{center}
\includegraphics[width=85mm,angle=0]{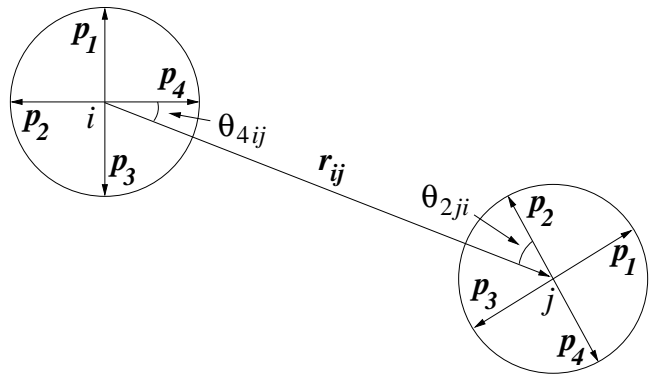}
\caption{\label{fig_potential} A schematic representation of the
	geometry of the interaction between two particles.
	 For clarity, we depict a two-dimensional analogue
	of the three-dimensional model used in this work.
	In this two-dimensional model, the particles have four patches arranged regularly
	with their directions described by the patch vectors, $p_i$. In the 
	particular case shown in the figure, patch 4
	on particle $i$ interacts with patch 2 in particle $j$ because
	they are the closest to the interparticle vector.}
\end{center}
\end{figure}

Our model consists of spherical particles with a given number of 
attractive patches whose geometry
is specified by a set of patch vectors. The total potential
can be written as a sum of two-body terms that depend on the
distance between two particles $r_{ij}$, but also on
their relative orientations $\mathbf{\Omega}_i$ and $\mathbf{\Omega}_j$:
\begin{equation}
U ({\bf r}_1, ..., {\bf r}_n, {\bf \Omega }_1, ...,  {\bf \Omega }_n) 
=  \sum_{i=1}^{N-1} \sum_{j=i+1}^{N} V(r_{ij},{\bf \Omega}_i,{\bf \Omega}_j)
\end{equation}
The interaction between two particles is described by a
potential with an isotropic repulsive core and an angular-dependent
attractive term:
\begin{equation}
V({\bf r}_{ij},{\bf \Omega}_i,{\bf \Omega}_j)  = 
\begin{cases}
V_{LJ}(r_{ij})  & r_{ij} <  \sigma_{LJ} \\
V_{LJ}(r_{ij})V_{ang}(\widehat{\mathbf{r}}_{ij},{\bf \Omega}_i,{\bf \Omega}_j) & r_{ij} \ge  \sigma_{LJ} 
\end{cases}  
\label{eq2}
\end{equation}
where $V_{LJ}(r)$ is the Lennard-Jones potential:
\begin{equation}
V_{LJ} (r)= 4 \epsilon \left[ \left( \frac{\sigma_{LJ}}{r}\right)^{12} -
          \left( \frac{\sigma_{LJ} }{r} \right)^{6} \right]
\end{equation}
$\epsilon$ is the pair well depth and $V_{LJ}(\sigma_{LJ})=0$.
Additionally, for computational efficiency, the potential is truncated and 
shifted using a cutoff distance of $2.5\,\sigma_{LJ}$. 
The attractive interaction is modulated by a product of
Gaussian functions that are centred at the position of each patch: 
\begin{equation}
V_{ang}(\widehat{r}_{ij},{\bf \Omega}_i,{\bf \Omega}_j)= \exp \left(-\frac{\theta_{k_{min},ij}^2}{2\sigma^2 } \right)
         \exp \left(-\frac{\theta_{l_{min},ji}^2}{2\sigma^2 } \right) ,
\label{eqn4}
\end{equation}
where $\sigma$ is the standard deviation of the Gaussian,
$\theta_{k,ij}$ ($\theta_{l,ji}$) is 
the angle formed between patch $k$ ($l$) on atom $i$ ($j$) and 
the interparticle vector 
${\bf r}_{ij}$ ( ${\bf r}_{ji}$), and $k_{min}$ ($l_{min}$) is the patch
that minimizes the magnitude of this angle.
The interaction is a maximum when both patches are pointing at
each other along the interparticle vector ${\bf r}_{ij}$ and it will decrease
as the particles deviate further from this equilibrium orientation.
A schematic representation of two such interacting particles
is provided in Fig. \ref{fig_potential}.

The angular dependence of Eq.\ \ref{eqn4} mimics the orientational
dependence that exists in short range directional forces as is the
case for hydrogen bonding. The use of $\theta_{k_{min},ij}$ and 
$\theta_{l_{min},ji}$ means that for a given pair of particles only a single
patch on each particle is involved in the interaction, i.e. the possibility of 
`double hydrogen bonding' is removed by the use of Eq.\ \ref{eqn4}.

One of the advantages of this potential is its simplicity. It is easy 
to implement and it is computationally not very expensive to evaluate. 
Furthermore, the model includes the anisotropy in a very simple and
flexible way. Simply by changing the number and position of patches, each
of which is defined by a vector in the particle reference system, 
it is possible to obtain a wide range of anisotropic potentials.
Another advantage is that the well-characterized Lennard-Jones potential
can be obtained as a limiting case of the current model when the width of the 
patches becomes increasingly large. This feature can be particularly useful 
if one wants to study the effect of the anisotropy, going from
a very anisotropic model to the isotropic limit. 

We shall use reduced units throughout, so that $U^*=U/\epsilon $,
$p^*=p/(\epsilon /\sigma_{LJ} ^3)$,
$T^*=T/(\epsilon /k_B)$ and $\rho^* = \rho \sigma_{LJ}^3$. Consequently,
the only parameter that needs to be specified to fully characterize
the interaction between two particles is $\sigma $, 
i.e., the angular width of the attractive patches (see Eq. \ref{eqn4}).
In the present calculations, we use particles with six patches that have
an octahedral arrangement, and 
since we wish to study the model in a regime where there
is a stable low-density crystal, we have chosen to use a relatively 
narrow patch width, namely $\sigma =$0.3 radians. 
In previous work\cite{jon} it has been shown that, for this value
of $\sigma $ and at not very high pressures, a low density 
sc crystal is formed spontaneously from the
liquid at sufficiently low temperature. In particular, at a constant
pressure $p^*=$0.1 crystallization occurred at $T^*\approx$0.17.

%%%%%%%%%%%%%%%%%%%%%%%%%%%%%%%%%%%%%%%%%%%%%%%%%%%%%%%%%%%%%%%%%%%%%%%
\subsection{Solid structures}
%%%%%%%%%%%%%%%%%%%%%%%%%%%%%%%%%%%%%%%%%%%%%%%%%%%%%%%%%%%%%%%%%%%%%%%

There are several crystalline structures that one might expect to be stable
for particles with an octahedral arrangement of the patches. 
The most simple structure is a simple cubic crystal.
In this structure, each of the six patches
points directly at one of the six nearest neighbours, and none
of the interactions will be frustrated.
However, this crystal has a relatively low density, 
e.g.\ if the nearest-neighbour separation is equal to the minimum in
the pair potential, $\rho^*=1/\sqrt{2}=0.707$,
and so it is expected that new denser crystalline phases will appear when 
the system is exposed to moderate to high pressures.
	
\begin{figure}[!tb]
\begin{center}
\includegraphics[width=85mm,angle=0]{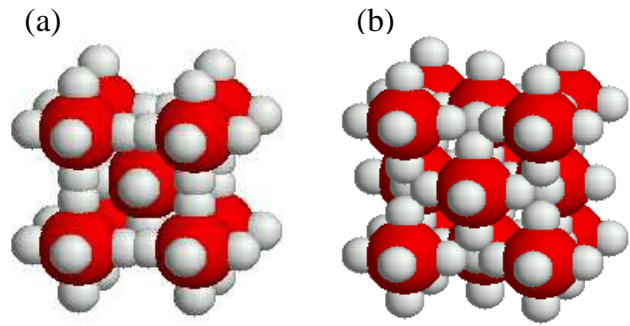}
\caption{\label{fig1} (Colour online) Unit cells of the 
	(a) bcc  and (b) orientationally ordered fcc-o structures.
        As mentioned in the text, in both cases the patches
	are aligned with the second neighbours.}
\end{center}
\end{figure}

Two possible higher-density crystals are the body-centred-cubic (bcc) and 
face-centred-cubic (fcc) lattices.
However, in these structures, due to the symmetry of the potential,
it is not possible to orient the particles
so that each patch is pointing directly towards one of its 
nearest neighbours. 
For a bcc crystal, it is possible to simultaneously align only
two of a particle's patches with its first neighbours.
However, an orientationally
ordered structure can be formed when the six patches of a given particle
are pointing towards
its six second nearest neighbours (see Figure \ref{fig1}). 
This structure can also be viewed as two interpenetrating simple cubic 
lattices that, as we will show, almost do not interact with each other. 
As the patches form an angle of $\pi /4$ radians
with the first neighbours, there will 
almost be no interactions between first neighbours, except when
they are closer than the repulsive 
barrier ($\sigma_{LJ}$). Furthermore, when the nearest-neighbours separation 
is $\sigma_{LJ}$, the distance betweeen second nearest neighbours is 
$2\sigma_{LJ}/\sqrt{3}=1.1547\sigma_{LJ}$ which is only slightly longer
than the minimum in the pair potential that occurs at 
$2^{1/6}\sigma_{LJ}=1.1225\sigma_{LJ}$. Therefore, the energetic cost of 
interpenetrating the two lattices is relatively small, but as the density
is significantly higher ($\rho^*=1.299$ for the above geometry) the bcc lattice
will have a more favourable enthalpy at moderate pressures.
In some senses, this structure has similarities to the structures of 
ice VII and ice VIII.  Both consist of two interpenetrating cubic ice lattices
that do no have any interconnecting hydrogen bonds between 
them.\cite{icebook}

For the fcc lattice, it is also possible to generate an
orientationally-ordered structure, 
%as in the bcc lattice, 
where the six patches are pointing towards six of the second nearest 
neighbours (we refer to this structure, which is depicted in Figure 
\ref{fig1}, as fcc-o). However, it is significantly higher in energy than
the bcc lattice, because in this case the ratio of the second to 
the first nearest-neighbour distance is $\sqrt{2}$, and so if there is to
be no repulsion between nearest neighbours (e.g.\ if 
their separation is $\sigma_{LJ}$ and $\rho^*=\sqrt{2}$) 
then the second nearest neighbours are significantly 
further apart than the minimum in the pair potential.

It is also possible to generate an orientationally-ordered 
tetragonal crystal, in which the centres of mass of the
particles are disposed as in a fcc lattice, but where 
four of the patches point towards first neighbours and
the other two patches point towards second neighbours. This structure
has a body-centred tetragonal unit cell ($a = b \ne c$,
where $a$, $b$ and $c$ are the moduli of the lattice vectors),
as the cubic symmetry of the fcc lattice has been broken
due to the orientation of the particles
(see Figure 1 in Reference \onlinecite{martensitic}
for an explanation of how to get the tetragonal unit cell in 
an analogous structure for a model of oppositely charged colloids).
However, in $NpT$ simulations in which each of the edges
of the box was allowed to vary independently,
we found the tetragonal crystal to undergo a transformation 
to an orientationally ordered bcc lattice. 
This type of deformation has 
been previously observed in other systems, and it is an example of a 
martensitic transition.\cite{martensitic} 
The transformation occurs by a shortening of the $c$ edge, until
$c=a$ and cubic symmetry is recovered.
In view of these results, we have not further considered the tetragonal 
structure in our calculations. Nevertheless, one cannot exclude the 
possibility that there might be some range of pressure and temperature where
the tetragonal structure is thermodynamically stable.
%Further simulations would be needed to clarify this issue.

At sufficiently high temperatures when the attractive interactions become
negligible, a translational ordered, but orientationally
disordered fcc structure, i.e. a plastic phase, will also appear. 
We refer to this structure as fcc-d (or PC).

%%%%%%%%%%%%%%%%%%%%%%%%%%%%%%%%%%%%%%%%%%%%%%%%%%%%%%%%%%%%%%%%%%%%%%%
\subsection{Equation of state for the fluid and solid phases}
%%%%%%%%%%%%%%%%%%%%%%%%%%%%%%%%%%%%%%%%%%%%%%%%%%%%%%%%%%%%%%%%%%%%%%%

The equation of state for the solid and the liquid phases was calculated 
using Monte Carlo (MC) simulations in the $NpT$ ensemble. 
Between 10 and 20 simulations were performed to determine the equation
of state along each isotherm. 
Each simulation consisted of 40000 MC cycles, following
an equilibration period of the same length. A MC cycle
was defined as $N$ attempts to translate a particle, plus $N$ attempts to rotate
a particle and two attempts to change the volume of the simulation
box, $N$ being the number of particles in the system. 
During the equilibration period, the maximum 
translational and rotational displacements were
adjusted to obtain a 40\% acceptance probability and the maximum
volume change was chosen to obtain a 30\% acceptance probability.
In all the simulations, the output
configuration of a given state was used as the input for the following 
simulation. 
The number of particles used in our calculations was 216 for the 
sc phase, 250 for the bcc phase, 256 for
the fcc-o and fcc-d phases and 250 for the fluid phase. 
In all the cases, the length of the box was larger
than twice the cutoff in the potential.
Since all the solid structures that we considered were cubic, 
we performed $NpT$ simulations with isotropic scaling.

%%%%%%%%%%%%%%%%%%%%%%%%%%%%%%%%%%%%%%%%%%%%%%%%%%%%%%%%%%%%%%%%%%%%%%%
\subsection{Free energy calculations}
%%%%%%%%%%%%%%%%%%%%%%%%%%%%%%%%%%%%%%%%%%%%%%%%%%%%%%%%%%%%%%%%%%%%%%%

At at given temperature, coexistence between two phases occurs at the 
pressure $p$ where the chemical potential $\mu $ is the same for both
phases. Since $\mu /k_BT=G/Nk_BT=A/Nk_BT+pV/Nk_BT$
and the compressibility $z=pV/Nk_BT$ can be obtained via $NpT$ simulations,
a procedure to determine the Helmholtz free energy is needed.

The free energy of the fluid phase was estimated by integration from
a very low density state, where the fluid can be considered to
behave as an ideal gas:
\begin{equation}
\frac{A(\rho )}{Nk_BT} = \frac{A^{id}(\rho )}{Nk_BT} + \int_{0}^{\rho} 
	\frac{z(\rho ') - 1}{\rho '} d\rho '
\label{liquid}
\end{equation}
where $A^{id}(\rho )/Nk_BT = \ln(\rho \Lambda^3) - 1$. We took 
$\Lambda =\sigma_{LJ}$ because its value does not affect 
the coexistence properties.

	For the solid phases, we used the method described by Frenkel
and Ladd\cite{frenkel-ladd,frenkelbook}, as extended to non-spherical
potentials.\cite{frenkel_molphys1985,vegamonson,JCP_96_9060_1992,JCP_97_8543_1992,monson_jcp00} 
In this method, the free energy is
obtained by integration from the interacting Einstein crystal,
whose interactions are described by the Hamiltonian:
\begin{equation}
\begin{split}
& \frac{H (\lambda_1^*, \lambda_2^* )}{k_BT}  =  \frac{H_{0}}{k_BT}+ 
 \lambda_1^* \sum_{i=1}^{N} \left(\sin (\Psi _a )^2 + \sin (\Psi _b)^2 \right) \\ 
 & \hspace{1.7cm} + \lambda_2^* \sum_{i=1}^{N} ({\bf r}_i -{\bf r}_{0,i})^2
\label{eqnh}
\end{split}
\end{equation}
On the right hand side of this equation,
$H_{0}$ is the original potential and the second and third
terms are, respectively,
an orientational and a translational field that tend
to keep the particles at the positions and orientations 
of the perfect orientationally ordered lattice.
In the term for the orientational field,
$\Psi _a$ is the minimum angle formed by any of 
the vectors that define the position of the patches
in the particle's reference system
with respect to the $x$ axis of a fixed reference system
and $\Psi _b$ is the analogous quantity with
respect to the $y$ axis, where the fixed reference system 
has been chosen to be 
coincident with the orientation of the patches in the
perfect lattice. As the patch that defines
both angles must not be the same, when the same patch yields the minimum
angle with both the $x$ and $y$ axes, the patches that 
form the second least angles were also computed. 
The patches that will contribute to the orientational
field will be the pair that leads to the lowest energy for this term.
Note that the orientational field has been chosen 
so that it has the same symmetry as the particles.

In the term for the translational field, ${\bf r}_{0,i}$ is the lattice
position of particle $i$ (as given by its centre of mass).
$\lambda_1^* =\lambda_1/k_BT$ and 
$\lambda_2^* =\lambda_2/(k_BT\sigma_{LJ}^{2})$ are the coupling
parameters of the translational and orientational field, respectively. For
practical reasons, we chose to use $\lambda_1^*=\lambda_2^*=\lambda^*$, 
i.e. both fields are switched on simultaneously. With this choice,
the integral to the reference translational and orientational Einstein crystals can 
be performed at the same time, thus reducing the number of simulations
needed to perform the numerical integration to the reference system.
However, other choices of $\lambda_1$ and $\lambda_2$ are equally valid.
Therefore, by using the same coupling parameter for both
fields,
Eq. \ref{eqnh} can be simplified into the following form:
\begin{equation}
\begin{split}
& \frac{H (\lambda^*)}{k_BT} =  \frac{H_{0}}{k_BT}+ 
\lambda^* \sum_{i=1}^{N} \left(\sin (\Psi _a )^2 + \sin (\Psi _b)^2 \right)  \\
& \hspace{1.3cm} + \lambda^* \sum_{i=1}^{N} ({\bf r}_i -{\bf r}_{0,i})^2
\end{split}
\end{equation}

	The free energy difference between our model and the reference
Einstein crystal ($\Delta A_2$) is given by:
\begin{equation}
\frac{\Delta A_2}{Nk_BT}
 = \int_{\lambda^*_{max} }^{0}
	\left\langle \frac{\partial (H(\lambda^*)/Nk_BT)}{\partial \lambda^* }  \right\rangle_{\lambda^*}
  d\lambda^*
\label{eqna2}
\end{equation}
This integral was evaluated numerically using a Gauss-Legendre 
quadrature formula, with 10 or 20 points depending on the case
(a larger number of points is needed for the orientationally disordered
plastic phase).

The free energy of a 
non-interacting Einstein crystal is already known,\cite{frenkel-ladd} and the
free energy difference between an interacting and a non-interacting 
crystal ($\Delta A_1$) can be calculated numerically by averaging the interparticle 
energy over a simulation of the non-interacting Einstein crystal:\cite{frenkel-ladd,JCP_102_1361_1995}
\begin{equation}
\small
\begin{split}
& \hspace{1.5cm}\frac{\Delta A_1}{Nk_BT}= \frac{U_0}{Nk_BT} \\ 
& - \frac{1}{N} \ln \left\langle \exp \left[
            - \left( \sum_{i=1}^{N-1}\sum_{j=i+1}^{N} 
		\frac{V(r_{ij}, \Omega_i, \Omega_j)}{k_BT} - \frac{U_0}{k_BT} \right) \right] \right\rangle ,
\label{eqna1}
\end{split}
\end{equation}
where $U_0$ is energy of the perfect lattice.
The maximum value of $\lambda^*$ in Eq. \ref{eqna2} was chosen so that
the structure of the interacting Einstein crystal defined by Eq. \ref{eqnh}
was very similar to the perfect orientationally ordered lattice.

The free energy of the orientational field can
be estimated numerically by integrating its partition function over all
the orientations:\cite{vegamonson,sandler_jcp2004}
\begin{eqnarray}
 \frac{A^{orient}}{Nk_BT} = &
 - \ln \left(  \frac{1}{8\pi^2} \int
	\exp \left\{ - \lambda^* \left[ \sin(\Psi_a)^2 + \sin(\Psi_b)^2 \right] \right\} \times \right.  \nonumber \\
&        \left.   \sin(\alpha) d\alpha d\phi d\gamma \right)
\end{eqnarray}
where $\Psi_a$ and $\Psi_b$ depend on the three Euler angles, $\alpha$, 
$\phi$ and $\gamma$. This integral was evaluated numerically using the
Monte Carlo integration method and using at least $10^9$ points.

The final expression for the free energy of the solid is:
\begin{equation}
A^{tot}= A^{Einstein} + \Delta A_1 + \Delta A_2 + \Delta A_3 
\end{equation}
where $A^{Einstein}$ is the sum of the free energy of the translational
Einstein crystal\cite{frenkel-ladd} plus the free energy of the orientational
field $A^{orient}$. The term $\Delta A_3$ accounts for the fact that
the integration to the Einstein crystal was evaluated by
performing simulations in which the centre of mass of the system was fixed
(see Ref. \onlinecite{frenkel-ladd}). 

\begin{figure}[!t]
\begin{center}
\includegraphics[width=80mm]{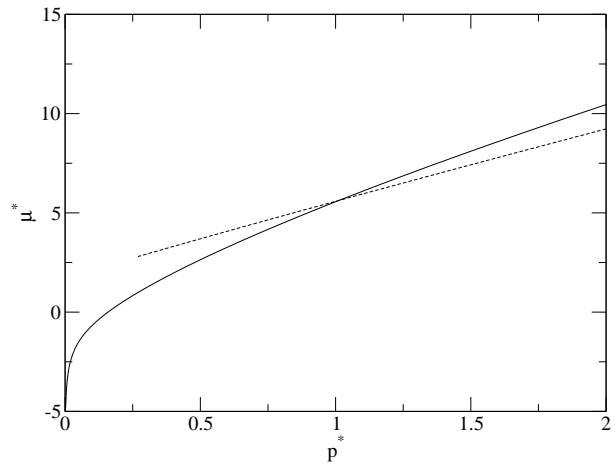}
\caption{\label{fig2} Determination of the coexistence point between the
	fluid (solid line)  and the bcc (dashed line)  phases at 
	$T^*=0.243$.}
\end{center}
\end{figure}

Once the free energy is known for a given state, the free 
energy along an isotherm can be obtained by integrating the equation
of state:
\begin{equation}
\frac{A( \rho_2) }{Nk_BT} = \frac{A(\rho_1)}{Nk_BT} + \frac{1}{k_BT} \int_{\rho_1}^{\rho _2} 
	\frac{p(\rho )}{\rho ^2} d\rho 
\label{integrosolid}
\end{equation}
The coexistence points are determined by imposing the conditions
of equal pressure and chemical potential, and can be obtained by
plotting the chemical potential against the pressure for both phases.
Coexistence is where the curves for the two phases cross, and 
Fig. \ref{fig2} shows such a plot for the bcc and the 
fluid phases at $T^*=0.243$.

%%%%%%%%%%%%%%%%%%%%%%%%%%%%%%%%%%%%%%%%%%%%%%%%%%%%%%%%%%%%%%%%%%%%%%%
\subsection{Coexistence lines}
%%%%%%%%%%%%%%%%%%%%%%%%%%%%%%%%%%%%%%%%%%%%%%%%%%%%%%%%%%%%%%%%%%%%%%%

The coexistence lines have been located using the Gibbs-Duhem integration
method introduced by Kofke.\cite{kofkeJCP,kofkeMP} 
In this method the coexistence line is
obtained by integration of the Clausius-Clapeyron equation:
\begin{equation}
\left( \frac{dp}{dT} \right) = \left( \frac{ \Delta u + p \Delta v}{T \Delta v} \right)
\label{clap}
\end{equation}
where $\Delta u$ and $\Delta v $ are the molar energy change and molar
volume change, respectively, between the two coexisting phases.

The Clausius-Clapeyron differential equation was solved
numerically with a fourth order Runge-Kutta algorithm, 
and using as the initial conditions the coexistence points obtained
with the thermodynamic integration method. The value
of the integrand in each of the four steps of the
Runge-Kutta method was evaluated by means of 
$NpT$ simulations that consisted of 10000 MC cycles, 
following 10000 cycles of equilibration.

%%%%%%%%%%%%%%%%%%%%%%%%%%%%%%%%%%%%%%%%%%%%%%%%%%%%%%%%%%%%%%%%%%%%%%%
\subsection{Direct coexistence simulations}
%%%%%%%%%%%%%%%%%%%%%%%%%%%%%%%%%%%%%%%%%%%%%%%%%%%%%%%%%%%%%%%%%%%%%%%

\begin{figure}[!tbh]
\begin{center}
\includegraphics[width=83mm,angle=0]{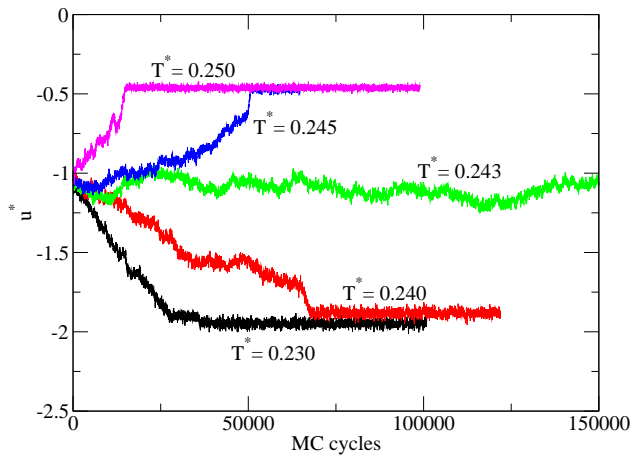}
\caption{\label{fig2b} (Colour online) The variation of the internal energy per particle  
	($u^*=U^*/N$) during $NpT$ simulations of a box containing the 
	bcc solid in contact with the fluid at $p^*=1.0$.
	A few trajectories at different temperatures are shown.}
\end{center}
\end{figure}

\begin{figure}[!tbh]
\begin{center}
\includegraphics[width=83mm,angle=0]{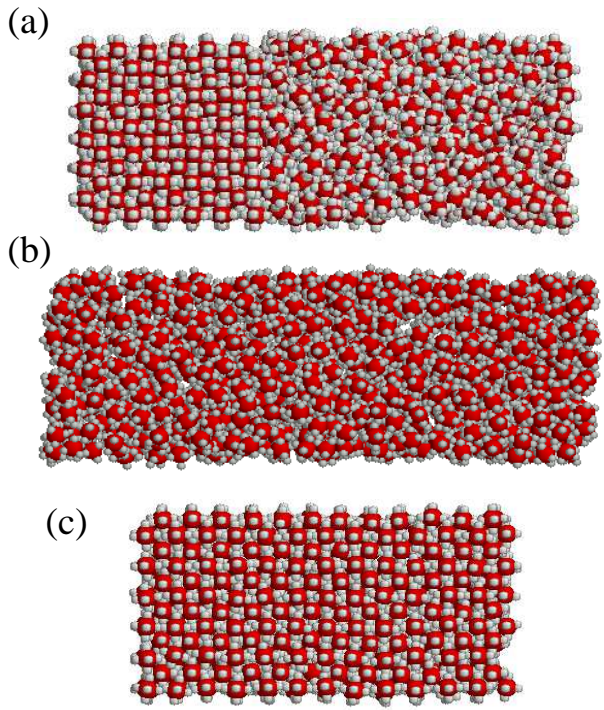}
\caption{\label{snapshot} (Colour online) (a) Snapshot of the initial 
	configuration of the simulation box containing the bcc and
	the fluid phases in contact. (b) and (c) Snapshots of the final
	configurations for $T^*=$0.245 and
	$T^*=$0.240, respectively. The pressure was set to $p^*=$0.1.}
\end{center}
\end{figure}

As a check of the above calculations, the melting points of the three solid 
phases were also estimated
using the direct coexistence method, first proposed
by Ladd and Woodcock.\cite{woodcock} In this work, we will follow the same
procedure as the one described in Ref. \onlinecite{ramon}.
A crystal with around 400-500 particles
was generated and subsequently equilibrated in the $NpT$-ensemble at a given 
pressure and temperature. This configuration was copied and heated to obtain a
liquid configuration, which was then equilibrated at the same conditions
as the solid but using a $NpT$-ensemble in which the volume of the
simulation box can only change by modifying one of the box lengths, which
we chose to be the box length along the $z$ direction. Therefore, both
the solid and liquid phases have the same periodic conditions 
along the $x$ and $y$ axes, and a solid-liquid interface can
be built by simply joining the liquid and solid configurations
along the $x-y$ plane.

	For a given pressure, the melting temperature can be estimated
performing $NpT$ MC simulations 
at different temperatures. If the temperature is above the melting
point, the solid phase will melt and the energy of the system
will increase. On the contrary, if the temperature is below
the melting, the solid phase will act as a nucleation seed and the
crystal will grow at the expense of the liquid, resulting in
a lowering of the system's energy. If it happens that the temperature is 
equal to the melting temperature, then the solid would be in equilibrium with 
the liquid, and the energy would remain constant. 
Using this procedure, we can determine a temperature interval 
which give upper and lower bounds for the melting temperature.

The results of $NpT$ simulations of a box containing the bcc solid
and liquid phases at $p^*=1.0$
are shown in Figure \ref{fig2b}. In this Figure, it can be seen that
the melting temperature
is approximately $T_{melt}^*=0.243$. At this temperature
the internal energy oscillates around an average value. 
At temperatures higher
than $T^*=0.243$, the energy increases until it reaches a constant
value, that corresponds to the situation when all the solid has melted.
Finally, at temperatures lower than $T^*=0.243$, the energy
decreases, again until it reaches a constant value, which, in this case,
means that all the fluid has frozen.
It is worth noting that both the time 
for the fluid to freeze and the time for the solid to melt
become shorter as the temperature of the system moves further away
from the melting temperature.
Figure 	\ref{snapshot} shows the initial configuration of the
simulation box, as well as two final states, one above and one
below the melting. 
A visual inspection shows that the final configuration at
$T^*=$0.245  corresponds to an homogeneous fluid phase
(Fig. \ref{snapshot} (b))
and, on the contrary, at $T^*=$0.240 it corresponds to a perfect
crystal (Fig. \ref{snapshot} (c)).

%%%%%%%%%%%%%%%%%%%%%%%%%%%%%%%%%%%%%%%%%%%%%%%%%%%%%%%%%%%%%%%%%%%%%%
%%%%%%%%%%%%%%%%%%%%%%%%%%%%%%%%%%%%%%%%%%%%%%%%%%%%%%%%%%%%%%%%%%%%%%
\section{Results}
\label{results}
%%%%%%%%%%%%%%%%%%%%%%%%%%%%%%%%%%%%%%%%%%%%%%%%%%%%%%%%%%%%%%%%%%%%%%

\begin{table*}[!bth]
\small
\centering
\begin{tabular}{cccccccc}
\hline\hline
     $T^*$  & $a_0$ & $a_1$ & $a_2$ & $a_3$ & $a_4$ & $a_5$ & $a_6$ \\
\hline
     0.200 & -0.6133262  &  2.3616349  &  9.5903228  & -55.187794 & 138.527889 & -122.51950   & -399.33624 \\
     0.243 &  0.1538809  &  2.6642095  & -2.2559535  &  2.8910917 & 29.6712406 & -39.785495   &  18.816227 \\
     0.500 &  1.2862246  &  1.6415799  &  8.4391218  & -30.150196 &  74.267261 &  72.615067   &  28.555569 \\
     3.00  &  1.6574717  &  2.3183855  & -1.5757928  &  10.982981 & -14.129931 &  10.595277   & -2.4231455 \\
 \hline\hline
\end{tabular}
\caption{\label{tbl2} Coefficients obtained by fitting the integrand of 
	Equation (\ref{liquid}) to a polynomial of degree six (Eq.\ (\ref{eq:fit})).}
\end{table*}

	Let us start by presenting the results for the fluid phase.
The free energy of the fluid phase was obtained by
thermodynamic integration from the very low density
limit, where the fluid can be considered to behave as
an ideal gas, to high densities (Equation \ref{liquid}). 
The integrand of this equation was evaluated by $NpT$ simulations
for different densities, and the results were
fitted to a polynomial of degree six:
\begin{equation}
\begin{split}
& \frac{z(\rho^*)-1}{\rho^*} = a_0 + a_1\rho^* + a_2\rho^{*2} + a_3\rho^{*3}
            +  a_4\rho^{*4}  \\
& \hspace{2cm}+ a_5\rho^{*5} + a_6\rho^{*6}
\label{eq:fit}
\end{split}
\end{equation}

\begin{figure}[!tbh]
\begin{center}
\includegraphics[width=83mm]{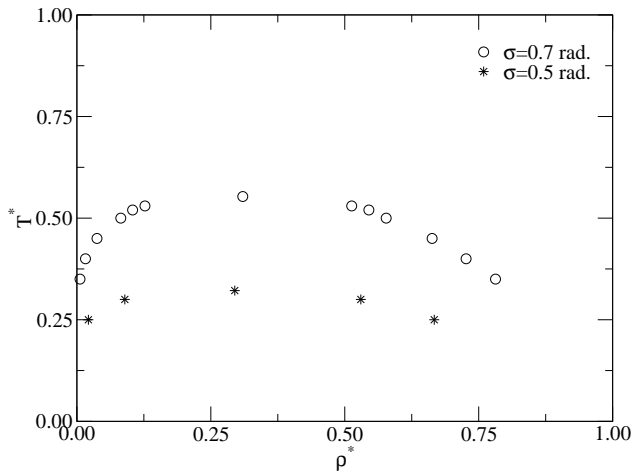}
\caption{\label{gibbs} Dependence of the vapour-liquid coexistence 
        curve on the patch width $\sigma $. At lower $\sigma$ we were unable to
	find the (metastable) coexistence curves.} 
\end{center}
\end{figure}

 	The coefficients resulting from
this fit for different 
isotherms are shown in Table \ref{tbl2}.
We have found no evidence of a vapour-liquid transition in any of
the isotherms studied, even at the lowest temperature we studied, $T^*=0.200$.
The coefficient $a_0$ provides an estimation of the second
virial coefficient of the model 
($B_2 = \lim_{\rho \to 0}  \frac{z-1}{\rho }$). As it can be seen,
the Boyle temperature, i.e., the temperature for which the
second virial coefficient vanishes, is located between 0.200 and 0.243.
A more precise estimate can be obtained by determining $B_2$
from numerical integration. Using this procedure we have
computed $B_2$ at $T^*=$ 0.200, 0.220, 0.230, 0.240, 0.500,
obtaining $B_2/\sigma_{LJ}^3= -0.556, -0.132, 0.031, 0.172, 1.386$. 
From these results, we estimated that
$T_{Boyle}^* \approx 0.228$.
An approximation to the critical temperature ($T_c$)
can be obtained from the Boyle temperature.
It is known that for another associative model,
the primitive model of water, 
$T_{Boyle}/T_c \approx 1.24$.\cite{vegamonson,sciortino_pmw}
Even though it is not clear how the anisotropy of
the model will affect this quotient,
we can obtain a rough estimate of
the critical temperature. 
Taking $T_{Boyle}^* \approx 0.228$, 
$T_c^* \approx 0.18$.
This estimate
is consistent with the fact that the equation of
state of the fluid phase along the isotherm $T^*=0.200$ does not show
any indication of a vapour-liquid equilibrium. 

	The possibility of a vapour-liquid phase separation 
was further explored by means of Gibbs ensemble (GE)
simulations.\cite{panagiotopoulos,panagiotopoulos_molphys88,smit_molphys89} 
Our GE simulations consisted on $5\times 10^5$ MC cycles for equilibration
plus $10^6$ MC cycles for obtaining averages, where a MC cycle 
was defined as $N$ attempts to translate a particle, 
$N$ attempts to rotate a particle, two attempts to change 
the volume and about 200 attempts to exchange particles
between the two boxes.
We simulated a system with 256 particles at a 
constant total number density $\rho^*=0.3$.
Several patch widths ($\sigma =$0.3, 0.5 and 0.7 radians) were 
considered and we found that,
consistent with previous calculations,\cite{sear_jcp1999,frenkel_jcp2003} the
vapour-liquid coexistence curve is very sensitive to the 
width of the patches, and that the critical temperature,
which was estimated by fitting the GE results
to the law of rectilinear diameter and to the
critical exponent scaling law,\cite{guggenheim45,rowlinsonbook} 
decreases as the patches become narrower (see Fig. \ref{gibbs}).
For the particular case $\sigma =0.3$ radians, we were
unable to locate any vapour-liquid coexistence points,
because, in this case, the simulations need to be performed
at very low temperatures, where it is particularly difficult to
obtain well-converged results.
In any case, the observed trends 
indicate that the critical temperature
for $\sigma=0.3$ must be very low and will
fall below the fluid-sc coexistence curve, which is greater than 
$T^*=0.2$ up to fairly low densities. These results
suggest that if vapour-liquid phase separation does exist,
it must be metastable with respect to freezing.
This is a relevant finding, because several experimental
studies have shown that globular proteins
also exhibit a vapour-liquid phase separation
that is metastable with respect to 
solidification.\cite{PNAS_88_5660_1991,JCP_107_1953_1997} 
Moreover, ten Wolde and Frenkel have suggested 
that crystallization occurs more 
rapidly in the proximities of this metastable critical
point.\cite{tenwolde} In that sense, it would be 
interesting to determine what is the maximum
patch width for which the phase separation between
a low density fluid and a high density fluid is metastable.

\begin{table}[!b]
\footnotesize
\begin{tabular}{lccccccc}
\hline\hline
    Structure & $T^*$  & $\rho^* $ & $\lambda^*_{max}$ & A$_{tot}$ & $\Delta A_2$ & $\Delta A_1$ & $U_0$\\
\hline
  sc 	&  0.100    & 0.762  & 25000 & -13.417  & -12.644 & -29.269  & -29.352 \\
  bcc   &  0.100    & 1.230  & 20000 & -11.637  & -10.784 & -28.667  & -28.723 \\
  fcc-o	&  0.100    & 1.360  & 20000 & 4.355    & -11.689 & -12.455  & -12.508 \\
  sc 	&  0.200    & 0.763  & 20000 & -1.095   & -14.284 & -14.616  & -14.669 \\
  bcc	&  0.200    & 1.175  & 20000 & 0.340    & -13.754 & -13.719  & -13.740 \\
  bcc	&  0.200    & 1.210  & 20000 & 0.690    & -13.001 & -14.122  & -14.146 \\
  bcc	&  0.243    & 1.147  & 20000 & 2.055    & -14.754 & -11.004  & -11.020 \\
  fcc-d (PC) &  0.500    & 1.204  & 20000 & 5.452    & -20.326 & -1.971   & -1.971 \\
  fcc-d (PC) &  3.00     & 1.283  & 20000 & 5.488    & -21.890 & -0.372   & -0.373 \\
  fcc-d (PC) &  3.00     & 1.376  & 20000 & 6.515    & -20.941 & -0.295   & -0.313 \\
 \hline\hline
\end{tabular}
\caption{\label{tbl1} Free energy of the solid phases, as obtained by
thermodynamic integration from the Einstein crystal. 
The free energies ($A_{tot}$, $\Delta A_2$ and $\Delta A_1$) and the 
lattice energy ($U_0$) are given in units of $Nk_BT$.}
\end{table}

\begin{table*}[!t]
\small
\centering
\begin{tabular}{lcccccc}
\hline\hline
     Solid phase & $T^*$  & $b_0$ & $b_1$ & $b_2$ & $b_3$ &  $b_4$ \\
\hline
     sc    &  0.100   &  26.7846382 & -69.2645290  &  6.67906462 &  54.0334053 &             \\
     bcc   &  0.100   &  612391.025 & -2010382.65  &  2475295.23 & -1354791.73 & 278126.549  \\
     fcc-o  &  0.100   & -631690.341 &  1383537.04  & -1010190.80 &  245893.687 &             \\
     bcc   &  0.243  & -570.084541 &  1657.42401  & -1613.44221 &  525.947816 &             \\
     fcc-d (PC)  &  3.00   & -1410.40288 &  3664.08695  & -3179.10036 &  943.459700 &             \\
\hline\hline
\end{tabular}
\caption{\label{solidfit} Coefficients of the polynomial fit to the equation of state
	of the solid phases. The points were fitted to a third-degree polynomial,
	except for the bcc structure at $T^*=0.1$, for which 
	a fourth-degree polynomial significantly improves the fit.}
\end{table*}

	The free energies of all the solid phases at several
thermodynamic states are given in
Table \ref{tbl1}. In some cases, we computed the free energy
for two states on the same isotherm, as these can then be used 
to test the thermodynamic consistency of our results 
(i.e.\ the free energy difference between two states as calculated 
using the Einstein crystal must be the same as that obtained
from integration of the equation of state).
In particular, for the bcc phase at $T^*=0.2$, the difference in free
energy between the states at reduced densities 1.210 and
1.175 is 0.350$Nk_BT$ as calculated
using the Einstein crystal approach, which compares well with
the value obtained by integrating the equation of state (0.355$Nk_BT$).
For the plastic fcc-d phase at $T^*=3.0$,
the agreement is also good 
(the free energy difference between the states at reduced densities 
1.376 and 1.283 computed using the two methods is 1.027 $Nk_BT$ and 
1.034 $Nk_BT$).
This thermodynamic consistency provides positive confirmation of 
the reliability of the calculations.

The free energy of the solid phases can be obtained at any other point on the
isotherm using Eq. \ref{integrosolid}. The equations of state ($p^*(\rho^*)$)
for the solid phases were obtained by performing $NpT$ simulations at 
different thermodynamic states along a given isotherm and 
fitting the results of the simulations to a polynomial of the form:
\begin{equation}
p^*(\rho^*) = b_0 + b_1 \rho^* + b_2 \rho^{*2} + b_3 \rho^{*3} + b_4 \rho^{*4}.
\end{equation}
The resulting data was fitted to a polynomial of degree three, except for
the bcc structure at $T^*=0.1$, for which a fourth-degree polynomial
lead to a much better fit.
The coefficients resulting from this fitting procedure are shown in 
Table \ref{solidfit}.

\begin{table}[!b]
\small
\centering
\begin{tabular}{llccccc}
\hline\hline
     Phase 1  & Phase 2  & $T^*$ & $p^*$ & $\rho^*_1$ & $\rho^*_2$ & \\
\hline
     fluid &   fcc-d (PC) & 3.00     &  48.0  & 1.182  &  1.270 & *\\
     fluid   &   fcc-d (PC) & 0.500    &  5.55  & 0.975  &  1.070 & \\
     fluid &   bcc   & 0.243    &  1.013 & 0.767  &  1.107 & * \\
     fluid   &   bcc   & 0.200    &  0.339 & 0.615  &  1.123 & \\
     fluid   &   sc    & 0.200    &  0.048 & 0.228  &  0.672 & \\
     sc      &   bcc   & 0.200    &  0.709 & 0.711  &  1.136 & \\
     sc    &   bcc   & 0.100    &  0.444 & 0.716  &  1.208 & * \\
     bcc   &   fcc-o  & 0.100    &  26.66 & 1.119  &  1.407 & * \\
 \hline\hline
\end{tabular}
\caption{\label{tbl3} Coexistence points obtained using the 
	thermodynamic integration method. The points marked with
	an asterisk were used as the starting points 
	for the Gibbs-Duhem approach. The other points 
	served to test our calculations.}
\end{table}

	Using these free energies, we calculated the 
coexistence points between all the phases that can be at equilibrium,
and the results are shown in Table \ref{tbl3}. 
In some cases, we have calculated
the coexistence between two phases at two different temperatures,
in order to verify that Gibbs-Duhem integration 
was able to give accurate results even for regions quite far from
the starting point.

\begin{table}[!b]
\footnotesize
\begin{tabular}{lccccc}
\hline\hline
& \multicolumn{2}{c} {Direct coexistence} & \multicolumn{2}{c} {Free energy calculations} \\
\cline{2-3} \cline{5-6}
    Solid & $p^*$  & $T^*$ &  & $p^*$  &  $T^*$ \\
\hline
  sc   &  0.758 & 0.232 $\pm $ 0.004 & & 0.758  &  0.229 \\
  bcc  &  1.000 & 0.243 $\pm $ 0.004 & & 1.013  &  0.243 \\
  fcc-d (PC) &  48.0  & 2.97  $\pm $ 0.02 &  & 48.0   &  3.00  \\
 \hline\hline
\end{tabular}
\caption{\label{tbl4} Melting points obtained using the direct coexistence
	method. For comparison, the coexistence points obtained from
	free energy calculations (see table \ref{tbl3}) are also shown.
       	Note that the sc-fluid coexistence point was obtained using the
	Gibbs-Duhem method (see table \ref{tbl5}).}
\end{table}

	As a further test of our calculations, the melting 
points of the solids have also been calculated using
the direct coexistence method. 
The melting points for all the solid phases obtained with this technique 
are shown in Table \ref{tbl4}.
The agreement between both methods is fairly good, the differences
being of the order of 1\%.

\begin{table}[!t]
\footnotesize
\begin{tabular}{llcccccc}
\hline\hline
    Phase 1 & Phase 2 & $p^*$  & $T^*$ & $\rho^*_1$ & $u^*_1$ & $\rho^*_2$ & $u^*_2$   \\
\hline
  fluid  &  fcc-d (PC) &   48.00   &   2.500   &   1.149 & 3.901    &    1.237  & 3.198 \\
  fluid  &  fcc-d (PC) &  38.19   &   2.000   &   1.112 & 1.913    &    1.207  & 1.485 \\
  fluid  &  fcc-d (PC) &   12.65   &   1.000   &   1.031 & 0.354    & 1.126 & 0.170  \\
  fluid  &  fcc-d (PC) &   0.624   &   0.500   &   -0.206 & -0.482  &   1.057 & -0.303 \\
\hline
  bcc  &  fcc-d (PC)    &    3.42   &   0.336   &   1.125 & -1.632  &   1.045 &  -0.470 \\
  bcc  &  fcc-d (PC)    &    5.00   &   0.355   &   1.156 & -1.580  &   1.121 &  -0.417 \\
  bcc  &  fcc-d (PC)    &    7.50   &   0.363   &   1.194 & -1.648  &   1.195 &  -0.353 \\
\hline
  bcc  &  fcc-d (PC)     &    10.25  &   0.365   &   1.224 & -1.678  &   1.249 &  -0.261 \\
  bcc  &  fcc-d (PC)     &    15.58  &   0.325   &   1.265 & -1.805  &   1.322 &  -0.119 \\
  bcc  &  fcc-d (PC)     &    22.36  &   0.250   &   1.294 & -1.949  &   1.374 &  -0.048 \\
  bcc  &  fcc-o     &    26.59  &   0.040   &   1.303 & -1.648  &   1.410 &  -0.353 \\
\hline
  fluid  &  bcc  &    2.915  &   0.320   &   0.927 & -0.390  &   1.121 &  -1.640 \\
  fluid  &  bcc  &    2.074  &   0.290   &   0.873 & -0.422  &   1.113 &  -1.759 \\
  fluid  &  bcc  &    0.778  &   0.230   &   0.729 & -0.494  &   1.110 &  -1.964 \\
\hline
  sc  &  bcc      &    0.751  &   0.225   &   0.709 & -2.154  &   1.113 &  -2.004 \\
  sc  &  bcc      &    0.582  &   0.150   &   0.714 & -2.493  &   1.173 &  -2.381 \\
  sc  &  bcc      &    0.382  &   0.080   &   0.716 & -2.746  &   1.222 &  -2.677 \\
\hline
  fluid  &  sc   &    0.758  &   0.229   &   0.728 & -0.494  &   0.709 &  -2.133 \\
  fluid  &  sc   &    0.200  &   0.218   &   0.486 & -0.378  &   0.677 &  -2.165 \\
  fluid  &  sc   &    0.010  &   0.181   &   0.059 & -0.070  &   0.675 &  -2.333 \\
 \hline\hline
\end{tabular}
\caption{\label{tbl5} Some of the coexistence points obtained with the Gibbs-Duhem
	method.}
\end{table}

	Using the coexistence points in Table \ref{tbl3} as
initial conditions, we traced the coexistence curves with the
Gibbs-Duhem method. Although we usually integrate the Clapeyron
equation, as given by Eq. \ref{clap}, we sometimes found it more 
convenient to integrate the equation $dT/dp=(T\Delta v)/(\Delta u + p\Delta v)$.
Some of the points obtained with the Gibbs-Duhem are shown in Table \ref{tbl5}.

\begin{figure}[!t]
\begin{center}
\includegraphics[width=83mm,angle=0]{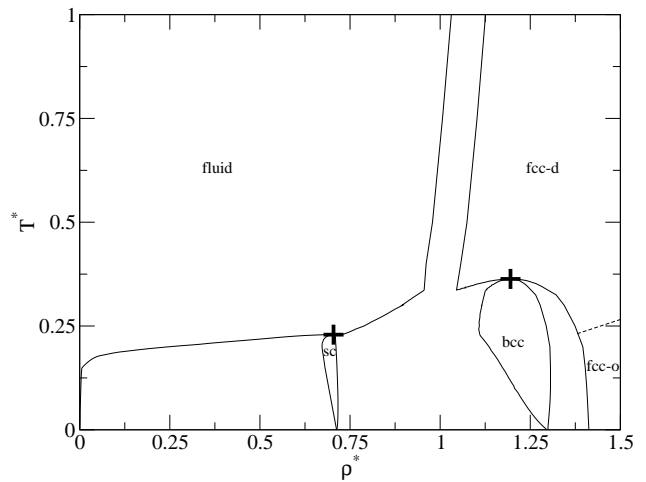}
\caption{\label{fig3} $T-\rho $ phase diagram of our octahedral six-patch particle
	system (with $\sigma =$0.3).
	Labels show the region of stability of each phase. The points 
	at which the reentrant behaviour occurs are indicated with a cross.}
\end{center}
\end{figure}

\begin{figure}[!t]
\begin{center}
\includegraphics[width=83mm,angle=0]{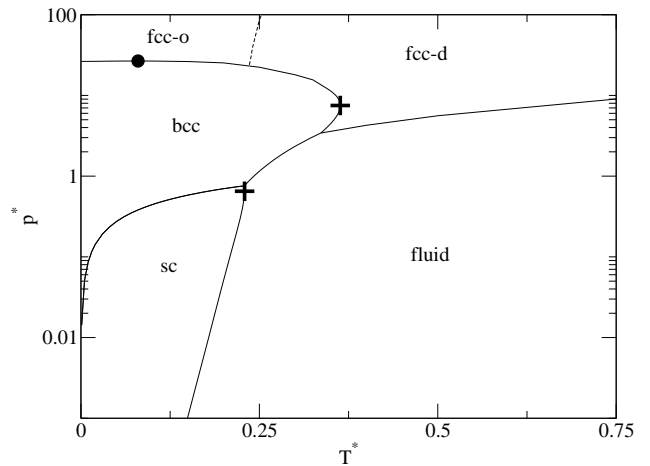}
\caption{\label{fig4} $p-T$ phase diagram of the octahedral six-patch particle
	system (with $\sigma =$0.3).
	Labels show the region of stability of each phase. The reentrant
	behaviour occurs at the points at which there is a change of sign
	in the slope of the phase boundaries. These points are indicated by a 
        cross. The black circle shows the point where inverse melting occurs.}
\end{center}
\end{figure}

The $T-\rho $ and $p-T$ phase diagrams are shown in
Figures \ref{fig3} and \ref{fig4}, respectively.
The dashed line in the diagrams shows a transition 
between the orientationally-ordered
and disordered fcc structures. These points have been estimated by
heating the ordered fcc-o structure and monitoring the internal energy.
This quantity exhibits a quite abrupt change when orientational order is lost,
and the transition temperature was chosen as the temperature where 
the internal energy curve shows an inflection point.

	As expected, the phase diagram shows multiple solid phases.
Firstly, at high temperature, where the behaviour is dominated by the 
repulsions between the particles, the fluid freezes into a plastic
crystal phase, the fcc-d phase (i.e., a lattice with fcc structure with respect 
to the centre of mass, but with orientational disorder). 
Secondly, at intermediate temperature,
it freezes into a bcc structure. 
Finally, at low temperature, 
it freezes into a low density sc solid.
The sc structure is only stable at fairly low pressures,
as it is possible to obtain a more dense phase, the bcc crystal, just by
introducing a single atom at the centre of the unit cell without
a large energetic penalty (see the discussion above and Table \ref{tbl1}). 
The bcc phase remains stable up to considerably higher
pressures. As the bcc structure is energetically much more favourable than
the fcc-o structure (Table \ref{tbl1}), the bcc crystal is only destabilized 
at densities for which the first neighbours are at
distances close to the LJ repulsive core $\sigma_{LJ}$.
In the fcc-o structure, the patches
are pointing to the second neighbours, which are at
a distance considerably larger than the LJ minimum (approximately 
$\sqrt{2}\sigma_{LJ}$ or larger). 

	It is worth noting that the sc, bcc and fcc-o phases are all stable
at $T^*=0$. At this temperature, the sc and bcc structures are both stable 
over a finite, but very small range of density, because under these 
circumstances the solid becomes almost incompressible.

\begin{figure}[!tbh]
\begin{center}
\includegraphics[width=83mm]{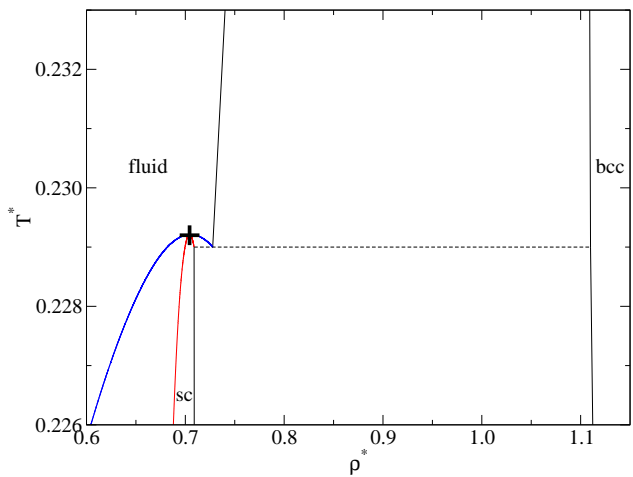}
\caption{\label{fig5} (Colour online) Detailed view of the phase diagram
	in the region of the sc-liquid-bcc triple point. The triple point
	is shown with a dashed line. 
        Labels indicate the region of stability of each phase. 
	The point at which reentrant behaviour occurs is indicated by a cross.}
\end{center}
\end{figure}

\begin{figure}[!tbh]
\begin{center}
\includegraphics[width=83mm]{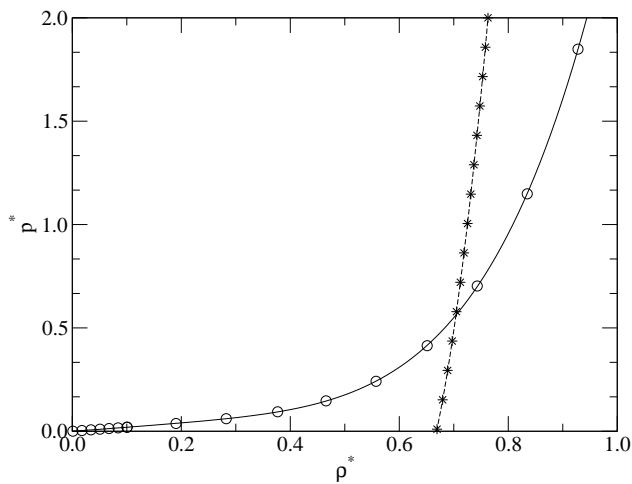}
\caption{\label{fig6} Equation of state for the liquid phase (circles)
	and the sc crystal (asterisks)
	at $T^*=0.2$.}
\end{center}
\end{figure}

	The phase diagram exhibits at least two triple points whose thermodynamic
states are given in table \ref{tbl6} (we have not studied in detail 
the triple point where the bcc, fcc-d and fcc-o phases coexist). 
At one of these triple points, the sc crystal coexists both with the liquid
and the bcc solid phases. This triple point
is somewhat unusual  and a magnified view of this region of the phase diagram 
is shown in Figure \ref{fig5}.
At the triple point, the sc crystal shows a slightly lower density than the 
liquid, the bcc crystal being the denser phase. 
The sc crystal is thermodynamically stable in a very narrow
window of temperatures above the triple point, and in this region
the sc-liquid coexistence lines adopt a dome-like shape. 
At the top of the dome, the liquid and the sc crystal are in equilibrium, 
both with the same density. The transition is first order though, 
because the enthalpy difference between both phases is not zero. 
For a small range of temperatures below this point, there are two 
values of the pressure at which the liquid and sc phases are in equilibrium. 
%for the same temperature. 
At the lower coexistence pressure, the sc crystal is the more dense phase.
However, the situation is reversed at the higher coexistence pressure,
where the liquid is in equilibrium with a lower density sc phase.
This means that the coexistence curve shows a reentrant behaviour in the 
vicinity of the triple point, i.e.\ there is a change in 
the sign of the slope of $dp/dT$ (Fig.\ \ref{fig4}) at the value of 
the pressure that coincides with the top of the dome in the $T-\rho$ diagram 
(Fig.\ \ref{fig5}). 
This kind of reentrant behaviour,
first speculated by Tammann to occur for water (see Ref. \cite{tamman}), has been
found by computer simulations both for primitive models of 
water\cite{vegamonson} and for realistic models of water.\cite{vegaprl04} 
The origin of this
behaviour can be seen in Figure \ref{fig6}. Since the compressibility
of the solid is very small, the fluid can be less dense and more dense
than the solid at different points along an isotherm. 
This leads to the existence of
two coexistence pressures for a given temperature (one low and one high),
which results in a reentrant behaviour. The low compressibility
of the solid is related to the strong directionality of the 
patchy bonds in our model
(or hydrogen bonds in the case of the water models).

\begin{table}[!t]
\begin{tabular}{lllccccc}
\hline\hline
    Phase 1 & Phase 2 & Phase 3 & $T^*$  & $p^*$  & $\rho^*_1$ & $\rho^*_2$ & $\rho^*_3$ \\
\hline
  sc       & liquid  &  bcc  &  0.229  & 0.758 & 0.709 & 0.728  & 1.110  \\
  liquid   & fcc-d (PC)  &  bcc  &  0.336  & 3.42  & 0.956 & 1.045  & 1.125  \\
 \hline\hline
\end{tabular}
\caption{\label{tbl6} Thermodynamic states for the two triple points
in the phase diagram.} 
\end{table}

	At a higher temperature, there is another 
triple point, where the fluid, the fcc-d plastic
and the bcc crystal are at equilibrium. Curiously, the fcc-d plastic phase
can be less dense than the bcc crystal. 
As the plastic phase is favoured by its high entropy not energy 
(Table \ref{tbl1}), the density can change considerably without any accompanying
large changes in the energy (as long as the repulsive cores of the 
particles do not overlap).
By contrast, the bcc phase is stabilized by its low energy, 
and deviations of the second nearest neighbour distance from the 
minimum in the LJ potential will result in a significant energetic penalty.
As a consequence of the wider range of densities possible for the
disordered fcc-d phase, the coexistence curves again show a
reentrant behaviour and adopt a dome-like shape in the $\rho-T$ phase diagram.

It is also worth mentioning that the bcc--fcc-o coexistence line 
seems to show inverse melting at very low temperatures
($T^*\approx 0.08$, see black circle in Fig. \ref{fig4}). This 
inverse melting is different from the reentrant melting
mentioned above in the sense that inverse melting is not caused by the 
fact that both phases have the same density (volume),
but because they have the same enthalpy (see Eq. \ref{eqn4}).
Tammann also speculated about this possibility,\cite{tamman}
and this unusual inverse melting behaviour has previously been observed 
for $^3He$ and $^4He$ and for poly(4-methylpentene-1).\cite{rastogi,greer,stillinger}

The results of this work are consistent with previous calculations using
a somewhat similar six-patch model,\cite{sandler_jcp2004} 
albeit with differences due to the differences in the potentials used. 
Chang \emph{et al.} also found several solid phases, including the plastic 
fcc-d phase at high temperatures, an ordered fcc solid at moderate pressures
and low temperatures, and a sc solid at low temperatures and 
pressures.\cite{sandler_jcp2004} 

%%%%%%%%%%%%%%%%%%%%%%%%%%%%%%%%%%%%%%%%%%%%%%%%%%%%%%%%%%%%%%%%%%%%%%%
%%%%%%%%%%%%%%%%%%%%%%%%%%%%%%%%%%%%%%%%%%%%%%%%%%%%%%%%%%%%%%%%%%%%%%%
\section{Conclusions}
\label{conclusion}
%%%%%%%%%%%%%%%%%%%%%%%%%%%%%%%%%%%%%%%%%%%%%%%%%%%%%%%%%%%%%%%%%%%%%%%

We have computed the phase diagram of particles with six 
octahedrally-arranged patches for a case where the patches are
relatively narrow. Even for this relatively simple potential, a complex
phase diagram results with competition between multiple solid phases differing
both in their density and whether they are orientationally ordered, 
and with unusual features such as reentrance and inverse melting.

Consistent with previous results,\cite{sear_jcp1999,frenkel_jcp2003,sandler_jcp2004}
there is no stable gas-liquid phase separation in our model.
The lowest-energy structure is a simple cubic crystal, but this
structure is only stable at relatively low pressure. 
As the pressure increases, first a bcc and then an fcc phase become most 
stable, where both phases are orientationally ordered. However, on 
increasing the temperature, entropic effects become more important and 
a plastic fcc crystal becomes most stable.

Even though we have considered very simple anisotropic models, they are able to
predict the formation of low density crystals, in analogy with the
preference of the proteins to form open crystals.
In that sense, even though our model is still far from
real proteins, our work can potentially provide some insights into 
the complex fluid-solid equilibrium of proteins.
In future work, it will be interesting to explore how the geometry and
number of the patches will influence the phase diagram, and in particular
the structure of the stable crystalline phases. Particularly relevant to 
proteins may be the exploration of random, rather than just ordered, 
arrangements of the patches. 

Although our calculations have determined the region of thermodynamic
stability for each phase, it does not necessarily follow that these phases 
will be easily accessible from within these regions. 
Indeed there is increasing evidence that the dynamics of crystal nucleation
can depend sensitively on the nature of the crystalline phase
and also of the liquid,\cite{Fernandez03,Kelton04,Molinero06,Sear07} 
as has been seen in preliminary calculations for the current model.\cite{jon}
Therefore, in future work we are planning to study the nucleation 
dynamics for the current model, and in particular to explore 
how this dynamics depends on the geometry of the patches. Such information
might provide important insights that could help colloidal chemists in designing
their anisotropic particles to crystallize into the desired target structure.

\acknowledgements

This work was funded by grants FIS2004-06227-C02-02 of Direcci\'on 
General de Investigaci\'on and S-0505/ESP/0229 of
Comunidad Aut\'onoma de Madrid. 
We would also like to acknowledge financial help from projects 
MTKD-CT-2004-509249 from the European Union and from project
910570 from the Universidad Computense de Madrid. 
E.G.N. wishes to thank 
the Ministerio de Educaci\'on y Ciencia 
and the Universidad Complutense de Madrid for a Juan de la Cierva fellowship, 
and J.P.K.D. and A.A.L. are grateful to the Royal Society for financial support.
We would like to thank David J. Wales and Mark A. Miller for useful discussions, and
Emanuela Zaccarelli and Francesco Sciortino
for organizing the interesting workshop entitled: \emph{ Patchy Colloids, Proteins and 
Network Forming Liquids: Analogies and new insights from computer simulations},
held at the CECAM. We are also grateful to F. Sciortino for sending
us a copy of Reference \onlinecite{sciortino_pmw} prior to publication.

\end{document}